\documentstyle[12pt]{article}
\newcommand{\be}{\begin{equation}}
\newcommand{\ee}{\end{equation}}
\newcommand{\ba}{\begin{eqnarray}}
\newcommand{\ea}{\end{eqnarray}}
\newcommand{\bastar}{\begin{eqnarray*}}
\newcommand{\eastar}{\end{eqnarray*}}
\newcommand{\half}{{1 \over 2}}

\newcommand{\X}{{\cal X}}

\begin{document}
\begin{titlepage}
\begin{flushright}
UU-ITP 13/94 \\
HU-TFT-94-24 \\
hep-th/9407060 \\
\vskip 0.4cm
June, 1994
\end{flushright}

\vskip 1.0truecm

\begin{center}
{ \large \bf ON THE INFRARED LIMIT OF THE \\ \vskip 0.2cm
              CHERN-SIMONS-PROCA THEORY \\  }
\end{center}

\vskip 1.5cm

\begin{center}
{\bf Antti J. Niemi $^\star$ and V. V. Sreedhar $^\dagger$ } \\
\vskip 0.4cm
{\it Department of Theoretical Physics, Uppsala University
\\ P.O. Box 803, S-75108, Uppsala, Sweden \\}
\vskip 0.4cm
and \\
\vskip 0.4cm
{\it $^\star$ Research Institute for Theoretical Physics,
Helsinki University,
Siltavuorenpenger 20 C, FIN-00170 Helsinki, Finland  \\}
\vskip 0.0cm
\end{center}

\vskip 1.5cm

\rm
\noindent
We investigate a modification of the 2+1 dimensional abelian
Chern-Simons theory, obtained by adding a Proca mass term to
the gauge field. We are particularly interested in the infrared
limit, which can be described by two {\it a priori}
different "topological" quantum mechanical models.  We
apply methods of equivariant cohomology and the ensuing
supersymmetry to analyze the partition functions of these
quantum mechanical models. In particular, we find that a
previously discussed  phase-space reductive limiting procedure
which
relates these two models can be seen as
a direct consequence of our supersymmetry.

\vfill

\begin{flushleft}
\rule{5.1 in}{.007 in}\\
{\small E-mail: $~~$ {\small \bf
niemi@rhea.teorfys.uu.se} $~~~~$ {\bf
sreedhar@rhea.teorfys.uu.se} \\ }
{\small $^\dagger$ Supported by a World Laboratory Scholarship}
\end{flushleft}

\end{titlepage}

\vfill\eject

\baselineskip 0.65cm

The abelian and nonabelian versions of three-dimensional
Chern-Simons theory have found several applications both
in Physics  and Mathematics [1]. Here we discuss a simple,
gauge invariant modification of the abelian Chern-Simons theory,
obtained by  adding a Proca mass term [2] to the gauge field.
This modification introduces a new dimensionful parameter
which is a relevant perturbation of the Chern-Simons action
in the infrared (or long-distance) limit. Thus it
could have even physically relevant consequences,
for example in the quantum Hall effect or high-temperature
superconductivity where it could {\it e.g.} parametrize
finite size effects in an experimental set-up.

In the present Letter we shall apply loop space equivariant
cohomology and the ensuing
supersymmetry [3] to investigate the infrared limit of the abelian
Chern-Simons-Proca theory. In particular, we find that
this limit can be described by two {\it a priori}
different "topological" quantum mechanical models that have been
discussed previously in [4]. Furthermore,  we conclude
that the phase-space reductive  limiting  procedure which
was used in [4] to relate these two  quantum
mechanical models to each other, is a direct consequence
of a supersymmetry which emerges  when the corresponding
path integrals are formulated in terms of loop space
equivariant cohomology as described in [3].

The abelian Chern-Simons gauge theory
is defined by the action
\be
S_{CS} = - \int {B\over 2}\epsilon^{\mu\nu\lambda}A_\mu
\partial_\nu
A_\lambda
= \int {B\over 2} (A_1 \dot A_2 - A_2 \dot A_1) - B\cdot
A_0 (\partial_1
A_2 -
\partial_2 A_1)
\ee
where $B$ is a parameter. Using a Hamiltonian interpretation,
we can
view $A_1$ and $A_2$ as canonically conjugated
variables with the Poisson
bracket
\be
\{ A_i(x) , A_j(y) \} ~=~ - \frac{1}{B} \epsilon_{ij}\delta(x-y)
\ee
Alternatively, we can use the standard definition
\be
E_i ~=~ { \delta L \over \delta \dot A_i } ~=~ - \frac{B}{2}
\epsilon_{ij} A_j
\ee
of canonically conjugated momenta to conclude that (2) is the
Dirac bracket
that follows from the second class constraints
\be
E_i + \frac{B}{2} \epsilon_{ij} A_j ~ \approx ~0
\ee
In both interpretations, the time component $A_0$ is a Lagrange
multiplier for the first class constraint
\be
F_{12} ~=~ \partial_1 A_2 - \partial_2 A_1 ~\approx ~0
\ee
which defines the abelian gauge transformation that leaves (1)
invariant.
We can fix the gauge {\it e.g.} by using the Coulomb gauge
condition,
\be
\partial_i A_i ~=~ \partial_1 A_1 + \partial_2 A_2 ~\approx~0
\ee
With the bracket (2)  we then have
\be
\{ \ F_{12}(x) \ , \ \partial_i A_i(y) \ \} ~=~ \frac{1}{B}
\Box_x
\delta(x-y)
\ee
and as explained {\it e.g.} in [5] we may interpret the
Coulomb-gauge
Chern-Simons theory  as a second class constrained
system, with constraints
\ba
\phi_1 &=& F_{12} ~\approx~ 0 \\
\phi_2 &=& \partial_i A_i ~\approx~ 0
\ea
and
\be
\{ \phi_i (x) \ , \ \phi_j (y) \} ~=~ \frac{1}{B}
\epsilon_{ij} \Box_x \delta(x-y)
\ee
as the second-class constraint algebra.
The determinant of (10) is nonsingular when evaluated over the
non-zeromodes of the two dimensional Laplacian, and the path
integral
\be
Z ~=~\int [dA_1][dA_2] \ \delta(\phi_i) \ {\det} ' ||
\{ \phi_i (x) \ , \ \phi_j (y) \} || \ \exp\{ i \int {B\over 2}
(A_1 \dot A_2 - A_2
\dot A_1) \}
\ee
describes the quantum Chern-Simons theory in the Coulomb gauge.
We
recall [5] that a change in the gauge condition (6) corresponds
to a change of variables in this path integral.

Notice that in the action in (11) we do not have any
Hamiltonian, only a
kinetic term appears. Consistent with general properties
of constrained
systems, we can consider modifications where we add a
Hamiltonian $H$ which is consistent with the symmetries of
the theory, {\it i.e.}
the brackets $ \{ \phi_i \ , \ H \} \ $ vanish on the constraint
surface (8)-(9)
\be
\{ \phi_i \ , \ H \} ~\approx~0
\ee
An example of such a Hamiltonian is the Coulomb gauge
version of the
Proca mass term [2]
\be
H_P ~=~ \int \frac{m}{2}  A_i^2  ~=~
\int \frac{m}{2}  (A_1^2 + A_2^2)
\ee
This gives rise to the following modification
of the Chern-Simons action in (11),
\be
S_{CS} ~\to~  S_{P} ~=~ \int {B\over 2} (A_1 \dot A_2 -
A_2 \dot A_1) -
\frac{m}{2}  (A_1^2 + A_2^2)
\ee
Indeed, since
\be
\{\phi_1 \ , \ H_P \} ~=~ \{F_{12} \ , \ H_P  \} ~=~ -
\frac{m}{B}\partial_i A_i ~=~- \frac{m}{B} \phi_2 ~\approx~ 0
\ee
and
\be
\{\phi_2 \ , \ H_P \} ~=~ \{ \partial_i A_i \ , \  H_P \} ~=~
\frac{m}{B} F_{12}
{}~=~ \frac{m}{B} \phi_1 ~\approx~ 0
\ee
this action is gauge invariant in the
Coulomb gauge. This is fully consistent with the standard
result [2]
that the Proca mass term is a gauge invariant perturbation of
the Maxwell
action.

Instead of (14) we could also consider the following
modification of the Chern-Simons action,
\be
S_{CS} ~\to~ - \int   {B\over 2}\epsilon^{\mu\nu\lambda}
A_\mu\partial_\nu
A_\lambda + \frac{m}{2} A_\mu^2
\ee
which is manifestly Lorentz-invariant, and gauge invariant in the
Lorentz gauge
\be
\partial_{\mu} A_{\mu} ~=~0
\ee
Here we have the standard Proca mass term, and we refer  {\it
e.g.} to [2] for a discussion of its properties in the
context of standard
Maxwell theory. Here  we shall restrict ourselves to the
Proca mass
term in the Coulomb gauge Chern-Simons theory, since in
condensed matter
applications this is quite sufficient.

\vskip 0.3cm

The gauge field $A_\mu$ has dimensions $[A]\sim \frac{1}{L} $
in terms of a
length scale $L$. As a consequence the parameter $m$ in (14)
has the proper
dimensions of a mass. If we set $m = \frac{c}{L}$ where $c$ is a
dimensionless constant  we conclude that in the double scaling
limit $c\rightarrow \infty ,~ L\rightarrow \infty $ with $m$
constant, the Proca
mass term in (14) grows linearly with the length scale $L$.
Consequently it is a
{\it relevant} perturbation of the Chern-Simons action
in the long distance,
infrared limit. In particular, we can expect that the
Proca mass term could have
physically relevant consequences {\it e.g.} in condensed
matter applications of
the Chern-Simons theory such as quantum Hall effect and
high temperature
superconductivity.

Notice in particular, that the infrared behavior of the Proca
mass term is in contrast with the infrared behavior of
the conventional Maxwell
term $\frac{1}{\alpha}F_{ij}F_{ij}$ which is an
irrelevant perturbation of the
Chern-Simons action because in three dimensions it goes as
${1\over L}$ in the long distance limit.

In order to investigate the infrared properties of the
Coulomb
gauge Chern-Simons-Proca theory (14), (8)-(9)  we expand the
gauge
field $A_i$ in a Fourier series and   separate out the part which
plays an important role at long distances. This is the term which
is constant in space, and is denoted in the following by $q_i$,
\be A_i (x) = q_i (t) + \tilde {A}_i (x)
\ee
In the infrared limit the Chern-Simons-Proca action  then
reduces to
\be
S_P ~\to~ \int \left( {B\over 2}\epsilon_{ij} q_i\dot {q}_j
- {m\over 2}q_iq_i
\right)
\ee
We observe that this is precisely the topological
Chern-Simons quantum mechanical
model considered in [4].

{}From (20) we identify the Hamiltonian
\be
H = {m \over 2} q_i q_i
\ee
while the first term in (20) yields the bracket
\be
\lbrace q_i, q_j \rbrace = - {1\over B} \epsilon_{ij}
\ee
which we can again view either as a Poisson bracket in the
phase space with
coordinates $q_i$ or alternatively as a Dirac bracket
in a larger phase space
with canonical momenta
\be
p_i ~\equiv ~ { \partial L_P \over \partial \dot
q_i } ~=~ - \frac{B}{2}
\epsilon_{ij} q_j
\ee
and second class constraints
\be
p_i + \frac{B}{2} \epsilon_{ij} q_j ~\approx~ 0
\ee
In particular, if we use (24) to eliminate one of the
coordinates in (20), say
$q_2$, we conclude that (24) describes a  harmonic
oscillator with frequency
$\omega = \frac{2 m}{B}$, so that the corresponding
quantum mechanical partition
function is\footnotemark\footnotetext{Notice that
if we substitute (24) in (20),
we need to re-scale both $p$ and $q$ by a factor
of $\sqrt{2}$ in order to
properly normalize their commutator. This explains
the discrepancy of a factor 2
between our result and that in [4].}
\be
Z ~=~\exp \{ - \beta H_P \}
{}~=~ \half \cdot { 1 \over  \sinh \beta \frac{m}{B} }
\ee

As explained in [4], the spectrum of (20) can also
be described by a
topologically massive quantum mechanical model,
using a phase-space reductive
limiting procedure.  In the remaining of this Letter we shall
show how this result also follows  directly from the loop space
equivariant cohomology and the ensuing supersymmetry
developed in [3]: We first
construct a supersymmetric  version of (20)  that
contains the topologically
massive quantum  mechanical model of [4] in its
bosonic sector. The relation
in [4] between the spectra of the two theories is
then a simple consequence of
this supersymmetry.

In order to derive a supersymmetric version of (20),
we first consider a generic
coordinate system $z^a$ on a $2n$ dimensional phase
space $\Gamma $. In these
coordinates the Poisson (Dirac) brackets are given by
\be
\lbrace z^a, z^b \rbrace = \theta^{ab} (z)
\ee
which determines the components of the symplectic two form
$\theta$ on $\Gamma$,
\be
\theta = {1\over 2}\theta_{ab}~dz^a\wedge dz^b
\ee
In the present case we may for example identify $q^i$
with $z^a$ which
implies that
\be
\theta_{ab} = B\epsilon_{ab}
\ee
In these variables the canonical phase space path integral for the
Chern-Simons-Proca quantum mechanics (20) is
\be
Z = \int [dz^a]\ \sqrt{ \det ||  B \epsilon_{ab} || }  ~
\exp \{ i\int dt
\left({B\over 2}\epsilon_{ab}z^a \dot {z}^b - {m\over
2}z^az^a \right) \}
\ee
and if we introduce anticommuting variables $c^a$ we can
rewrite this as
\be
Z = \int [dz^a][dc^a]~ \exp i(S_B + S_F)
\ee
where
\be
S_B = \int dt \ \left({B\over 2}\epsilon_{ab}z^a\dot {z}^b - {
m\over
2}z^a z^a
\right)
\ee
and
\be
S_F = \int dt \ {B\over 2}c^a \epsilon_{ab}c^b
\ee

Following [3] we interpret (30) as a loop space
integral in a loop space $L\Gamma$, parametrized  by the
time evolution $z^a
\rightarrow z^a (t)$ with periodic boundary
conditions $z^a (t_i) = z^a (t_f)$.
In this loop space we define exterior derivative by lifting the
exterior derivative on the phase space $\Gamma $,
\be
d = \int dt~dz^a(t) {\delta\over\delta z^a (t)} \equiv
dz^a{\delta\over
\delta z^a}
\ee
The $dz^a (t)$ obtained by lifting
the basis of one forms $dz^a$ on   $\Gamma$ then
constitutes a basis of one forms
on the loop space $L\Gamma$, and we can
identify $dz^a$ with the anticommuting
variables $c^a(t)$.

We define a loop space Hamiltonian vector
field  $\X^a_P$, determined
by the bosonic part (31)  of the action
through the equation
\be
{\delta S_B \over\delta z^a(t)}=\Theta_{ab}(z; t,t^\prime
)\X^b_P (t^\prime)
\ee
where $\Theta_{ab} $ are the components of a loop space
symplectic
two form $\Theta $, obtained by lifting the components
of the symplectic two
form $\theta $ to $L\Gamma $ through the relation
\be
\Theta_{ab} (z; t, t^\prime) = \theta_{ab} (z)\delta (t-t^\prime )
\ee
In the present case this implies
\be
\Theta_{ab}(t,t^\prime ) = B\epsilon_{ab}\delta
(t-t^\prime )
\ee
so that the Hamiltonian vector field is
\be
\X^a_P = {\dot {z}^a\over 2} + {m\over B}\epsilon^{ab}z^b
\ee
Notice in particular, that the zeroes of (37) yield the
classical solutions of
the Chern-Simons-Proca theory.

Let $i_P$ denote loop space contraction along the
Hamiltonian vector field
$\X_P$,
\be
i_P = \X^a_P i_a
\ee
where the $i_a(t)$ form a basis of loop space
contractions which is dual
to $c^a(t)$,
\be
i_a(t)c^b(t^\prime ) = \delta_a^b(t-t^\prime )
\ee
We define a loop space equivariant exterior derivative by
\be
d_P = d + i_P
\ee
It gives the following loop space supersymmetry
transformation on the
variables $z^a$ and $c^a$,
\be
d_P z^a = c^a
\ee
\be
d_P c^a = \X^a_P
\ee
More precisely, if $L\Lambda$
is the DeRham complex on the loop space $L\Gamma $, $d_P$
maps the subspace of
even forms in $L\Lambda$ onto the subspace of odd forms
and vice versa.
Furthermore,
\be
d_P^2 ~=~ d i_P + i_P d ~=~ {\cal L}_P
\ee
is the loop space Lie derivative along the Hamiltonian
vector field
$\X_P$, and if we consider the invariant subspace
\be
L\Lambda_{inv} = \lbrace \xi\in L\Lambda \mid {\cal {L}}_P
\xi = 0
\rbrace
\ee
in this subspace $d_P$ is nilpotent and acts like an
exterior derivative.
In particular, we find that the action
$S_B + S_F$ is invariant under the supersymmetry
transformation (41)-(42)
\be
d_P (S_B+S_F) = 0
\ee
and consequently determines an element in the invariant
subspace (44).

If $g_{ab}$ is a metric on $L\Gamma $ which is Lie
derived by $\X_P $, the
function
\be
\psi =   g_{ab}\dot {z}^ac^b
\ee
is also Lie derived by $\X_P$,
\be
{\cal {L}}_P g = 0 ~ \Longrightarrow ~ {\cal {L}}_P \psi  = 0
\ee
In the present case, we can simply choose the flat
Euclidean metric $g_{ab}
= \delta_{ab}$. As explained in [3], if we now
add to the
action $S_B + S_F$ a $d_P$-exact piece of the form
\be
\lambda \cdot d_P\psi  = \lambda \delta_{ab}\lbrack \dot {
c}^ac^b + {1\over
2}\dot {z}^a
\dot {z}^b + {m\over B}\epsilon_{bc}\dot {z}^az^c \rbrack
\ee
where the parameter $\lambda $ has dimensions $[\lambda ]
\sim L $, the
corresponding path integral is independent of $\lambda$ and
coincides with the
original path integral (29).

Explicitly, this gives for our infrared limit of the
Chern-Simons-Proca action
(we now specialize to $z^a \to q^i$ and $c^a \to c^i$)
\be
S_P ~\to~ \int \lambda \delta_{ab}\dot {q}^i\dot {q}^j + {
B^\prime\over 2}
\epsilon_{ij}q^i \dot {q}^j - {m\over 2}\delta_{ij}q^i q^j +
\lambda
\delta_{ij} \dot{c}^i c^j + {B\over 2}c^i\epsilon_{ij}c^j
\ee
where
\be
{B^\prime\over 2} = \left( {B\over 2} - {\lambda m\over
B}\right)
\ee

{}From the functional form of (49) we conclude in the usual
manner that we can
introduce momenta $p_i$, $\pi_i$ which are canonically
conjugate to $q^i$,
$c^i$,
\be
p_i \equiv {\partial L\over \partial \dot {q}^i}= \lambda \dot
{q}^i
- {B^\prime \over 2}\epsilon_{ij}q^j
\ee
\be
\pi_i \equiv {\partial L\over \partial \dot {c}^i} = \lambda
c^i
\ee
In this way we conclude that the Hamiltonian of (49) is
\be
H = H_B + H_F = {1\over 2\lambda}\left[ p_i + {B^\prime\over
2}\epsilon_{ij}q^j\right]
\left[ p_i + {B^\prime\over 2}\epsilon_{ik}q^k\right] + {m\over
2}q^i q^i
-{B\over 2}c^i\epsilon_{ij}c^j
\ee
which is the desired supersymmetric extension of the
Chern-Simons-Proca
Hamiltonian  (21). In particular, we observe that the
bosonic part
\be
H_B = {1\over 2\lambda}\left[ p_i + {B^\prime\over
2}\epsilon_{aij}q^j\right]
\left[ p_i + {B^\prime\over 2}\epsilon_{ik}q^k\right] + {m\over
2}q^i q^i
\ee
of (53)  coincides with the topologically massive
Hamiltonian discussed in [4],
except for the shift proportional to $\lambda$
displayed in equation (50). This
shift, innocuous as it is in so far as the qualitative
features of the spectrum
are concerned, will play an important role when we
proceed to reproduce the
results in [4].   Indeed, according to general
arguments [3] which are based on
the supersymmetry (41)-(42), we expect that the
partition function of (53) will
be {\it independent} of $\lambda$, and coincides
with the partition function
(25) of the original Chern-Simons-Proca quantum
mechanical model (20). This
reproduces the results obtained in [4] using a
phase-space reductive limiting
procedure.

We  shall now proceed to explicitly verify that the partition
functions of (20) and (49) indeed coincide, {\it
independently} of $\lambda$.
For this we introduce the following  variables
\be
p_\pm = \left( { \omega_\pm\over 2\lambda\Omega} \right)^{1
\over 2}  p_1
\ \pm \ \left({\lambda\Omega\omega_\pm\over 2}\right)^{1
\over 2}q^2
\ee
\be
q_\pm = \left({\lambda\Omega\over 2\omega_\pm} \right)^{1\over
2}q^1
\ \mp \ \left({1\over 2\lambda\Omega\omega_\pm}\right)^{1
\over 2}p_2
\ee
with the nonvanishing commutators
\be
\lbrack q_+, p_+\rbrack ~=~ \lbrack q_-, p_-\rbrack ~=~ i
\ee
where
\ba
\Omega &=& \sqrt{ \left({B\over 2\lambda }\right)^2 + {m^2
\over B^2} }
\\
\omega_\pm &=& \Omega \pm {B\over 2\lambda } \mp {m\over  B}
\ea
In terms of these variables the Hamiltonian $H_B$ splits
into two decoupled one
dimensional oscillators with frequencies $\omega_\pm$.
In terms of creation and annihilation operators
\ba
{a}^\star_\pm &=& {1\over 2\omega_\pm }(p_\pm + i
\omega_\pm q_\pm ) \\
a_\pm &=& {1\over 2\omega_\pm }(p_\pm - i\omega_\pm q_\pm )
\ea
such that
\ba
\lbrack a_+, {a}^\star_+\rbrack &=& 1 \\
\lbrack a_-,  {a}^\star_-\rbrack &=& 1
\ea
with all other commutators being zero,
we get for $H_B$
\be
H_B = \omega_+ ({a}^\star_+a_ + + {1\over 2}) + \omega_-
({a}^\star_-a_- +
{1\over 2})
\ee

We now proceed to discuss the fermionic part $H_F$:
Recalling the definition
of the canonical momentum conjugate to $c^i$   from
equation (52) and imposing
the anticommutation relations
\be
\lbrace c^i, \pi_j \rbrace = i\delta^i_j
\ee
we get the Dirac brackets
\be
\lbrace c^i, c^j \rbrace = i {\delta^{ij}\over \lambda }
\ee
Consequently we can realize the $c^i$ in terms of Pauli
matrices $\sigma^1$ and
$\sigma^2$ through the relations
\ba
c^1 &=& \sqrt{ {i\over 2\lambda} } \ \sigma^1
\\
c^2 &=&
\sqrt{ {i\over 2\lambda} } \ \sigma^2
\ea
and we find that we can rewrite $H_F$ as
\be
H_F = {B\over 2\lambda }\sigma^3
\ee
and combining (64), (69) we get for the Hamiltonian (53),
\be
H  = \omega_+ (a_+^\star a_+ + {1\over 2}) + \omega_- (a_-^\star
a_- +
{1\over 2}) + {B\over 2\lambda }\sigma^3
\ee

Let
\be
\mid N_+, N_-, N_F \rangle  ~~~~~~~~~~~ N_\pm = 0, 1, 2,
\cdots ~; ~~~~~~~
N_F =  0, 1
\ee
denote the simultaneous eigenstates of the number operators
for our two bosonic
and one fermionic oscillators. The spectrum of the
Hamiltonian (70) is then given by
\be
E(N_+, N_-, N_F) = \left[\Omega - (-1)^{N_F}{B\over
2\lambda}\right] +
\omega_+N_+ + \omega_-N_-
\ee
and we can directly evaluate the  partition function
\be
Z ~\equiv~ Tr \{ (-1)^{N_F} e^{  -\beta H  } \}  = \sum_{N_+,
N_-, N_F}
(-1)^{N_F} \exp \{ -\beta E (N_+, N_-, N_F) \}
\ee
We find
\be
Z = \left[ \exp \{ -\beta \left( {B\over \lambda }
\right) \} -
\exp \{ -\beta \left( -{B\over \lambda } \right) \}
\right] \cdot
{\exp \{ - \half \beta \omega_+ \} \over 1 - \exp \{
-\beta\omega_+ \} }
\cdot { \exp \{ -\half \beta \omega_- \} \over 1 - \exp \{ -
\beta\omega_- \} }
\ee
and using (59) we get after  a little algebra
\be
Z = \half \cdot {1\over \sinh \beta {m\over B}}
\ee
which coincides with (25), as expected. In particular,
consistent
with our general arguments all $\lambda$-dependence
has disappeared
in (75). As a consequence we have reproduced
the results of [4]
directly, using supersymmetry which is based on
loop space equivariant
cohomology.  Notice in particular, that the
infinite subtraction that is
required in [4] for the zero point spectra of
the two theories to coincide, is
here taken care of by the supersymmetry.

\vskip 0.8cm

In conclusion, we have investigated a simple gauge invariant
modification of the three dimensional abelian
Chern-Simons theory,
obtained by adding the Proca mass term. This
modification
of the Chern-Simons theory survives in the
long-distance limit, and
consequently determines a relevant infrared
perturbation of the original
theory that could have experimental
consequences {\it e.g.} in
quantum Hall effect and high-temperature
superconductivity.
Furthermore, we have found that in this
limit our model reproduces the
topological quantum mechanical models
investigated in [4]. In particular, we
have derived the results in [4] directly,
using supersymmetry which is
determined by loop space equivariant
cohomology. Since this supersymmetry
emerges from the symplectic structure
of the theory, we expect that similar
techniques could yield interesting results
also in the three dimensional
context of our model.

\vskip 1.5cm
{\bf Acknowledgements:} V.V.S. thanks Olav Tirkkonen and Alexei
Rosly for discussions, and the World Laboratory, Switzerland for a
Scholarship.

\vfill\eject

{\bf References}

\vskip 0.8cm

\begin{enumerate}

\item  F. Wilczek, ed. {\it Fractional Statistics and Anyon
Superconductivity}  (World Scientific, Singapore) (1990);
E. Witten, Comm. Math. Phys. {\bf 121}, 351 (1989)

\item J.D. Jackson, {\it Classical Electrodynamics, 2nd edition}
(John Wiley and Sons, New York) (1975); C. Itzykson
and J.-B. Zuber,
{\it Quantum Field Theory} (McGraw-Hill, New York) (1980)

\item  M. Blau, E. Keski-Vakkuri and A.J. Niemi, Phys. Lett.
{\bf B246}, 92 (1990)

\item  G. Dunne, R. Jackiw and C. Trugenberger,
Phys. Rev. {\bf  D41}, 661
(1990)

\item L.D. Faddeev and A.A. Slavnov, {\it Gauge
Fields - Introduction to
Quantum Theory} (Benjamin/Cummings, Reading MA) (1980)

\end{enumerate}

\end{document}